\documentclass[preprint,aps]{revtex4}
\usepackage {graphicx}

\begin{document}

\def\be{\begin{equation}}
\def\ee{\end{equation}}

\title[]{Variational approach to the time-dependent Schr\"{o}dinger-Newton equations}
\author{Giovanni Manfredi}
\address{Institut de
Physique et Chimie des Mat\'{e}riaux, CNRS and Universit\'{e} de
Strasbourg, BP 43, F-67034 Strasbourg, France}
\author{Paul-Antoine Hervieux}
\address{Institut de
Physique et Chimie des Mat\'{e}riaux, CNRS and Universit\'{e} de
Strasbourg, BP 43, F-67034 Strasbourg, France}
\author{Fernando Haas}
\address{Departamento de F\'{\i}sica, Universidade Federal do Paran\'{a}, 81531-990, Curitiba, Paran\'{a}, Brazil}

\date{\today}

\begin{abstract}
Using a variational approach based on a Lagrangian formulation and Gaussian trial functions, we derive a simple dynamical system that captures the main features of the time-dependent Schr\"{o}dinger-Newton equations. With little analytical or numerical effort, the model furnishes information on the ground state density and energy eigenvalue, the linear frequencies, as well as the nonlinear long-time behaviour. Our results are in good agreement with those obtained through analytical estimates or numerical simulations of the full Schr\"{o}dinger-Newton equations.
\end{abstract}

\maketitle

\section{Introduction}\label{sec:intro}
In recent years, there has been a renewal of interest in the set of nonlinear equations known as the Schr\"{o}dinger-Newton (SN) equations. These consist of the ordinary Schr\"{o}dinger equation
\be
i\hbar \frac{\partial \Psi}{\partial t} = -\frac{\hbar^2}{2m} \Delta \Psi +m V({\mathbf r},t) \Psi,
\label{schrodinger}
\ee
where the gravitational potential $V({\mathbf r},t)$, in the Newtonian approximation, is obtained self-consistently from Poisson's equation
\be
\Delta V = 4\pi G m|\Psi|^2,
\label{poisson}
\ee
where $m$ is the mass of the system and $G$ is the gravitational constant.
The source term in Poisson's equation is provided by a matter density $\rho({\mathbf r},t) = m|\Psi|^2$ that is proportional to the probability density as given by the wavefunction $\Psi({\mathbf r},t)$. The resulting equations are therefore nonlinear.

SN-type equations have been proposed in various areas of physics and astrophysics. For instance, it has been suggested that gravitation, unlike other forces, may not be quantized at all \cite{carlip}. In that case, the stress-energy tensor $T_{\mu \nu}$ in Einstein's equations should be replaced by its quantum-mechanical average $\langle T_{\mu \nu} \rangle$. The SN equations can thus be viewed as the nonrelativistic ($c \to \infty$) and Newtonian ($G \to 0$) limit of the modified Einstein's equations $G_{\mu \nu} = (8\pi G/c^4) \langle T_{\mu \nu} \rangle$. More formally, Giulini and Gro\ss ardt \cite{giulini2012} recently showed that the SN equations can be derived in a WKB-like expansion in $1/c$ from the Einstein-Klein-Gordon and Einstein-Dirac system.

In another context, the SN equations have been proposed as a fundamental modification of the Schr\"{o}dinger equation due to gravitational effects. Penrose \cite{penrose96,penrose98} and Diosi \cite{diosi} postulated that gravity might be at the origin of the spontaneous collapse of the wavefunction and proposed the (stationary) SN equations as a possible candidate for an approximate description of such gravitationally-induced collapse.

Finally, the SN equations have been used in an astrophysical context to study self-gravitating objects such as boson stars \cite{schunk, chavanis} or to describe dark matter by means of a scalar field \cite{guzman}.

Whatever their present theoretical status and possible applications, the SN equations represent a minimal model in which nonrelativistic quantum mechanics is coupled self-consistently to Newtonian gravity. As such, they are worth investigating in some detail, both for their static and their dynamical properties.

Many theoretical results on the SN equations were obtained in the past, using either analytical or numerical approaches \cite{moroz}. For instance, the energy eigenvalues (all negative) have been determined numerically with good precision \cite{harrison} and some analytical estimates exist on the lower bound for the ground state energy \cite{tod}. The linear stability properties of the ground state were also investigated \cite{harrison}.

In the time-dependent and fully nonlinear regime, virtually all results are numerical, with few exceptions whose validity is restricted to short time scales \cite{giulini2011}. An unexpected result was published a few years ago by Salzman and Carlip \cite{salzman}. In numerical simulations of spherically symmetric systems, these authors observed that, for masses above a certain critical value, the wavefunction ``collapsed'' at the origin, at least within the accuracy of their simulations\footnote{One should not mistake this gravitational collapse (whether it is real or not), with the quantum collapse of the wavefunction during a measurement, which is a nonunitary process. The time-dependent SN equations are unitary and thus cannot describe any such process.}. The most astonishing feature of these results was that the critical mass was far smaller than what could be expected from simple order-of-magnitude calculations. However, more recent calculations \cite{giulini2011,vanmeter} disagree with the results of Salzman and Carlip and set the critical mass at a value that is several order of magnitudes larger and consistent with analytical estimates.

In this work, we revisit the SN equations using a Lagrangian variational method \cite{haas,manfredi_ep}. With this approach, one can arrive at a single ordinary differential equation that describes the evolution of the width of the mass density. This method reproduces all the main results on the ground state and linear dynamics derived previously. In addition, this approach is not restricted to linear theory and can be used to investigate nonlinear oscillations or the long-time dynamics.
Finally, the mathematical simplicity of the governing equation makes it easy to intuit at a glance the salient features of the solutions.

\section{Derivation of the model}\label{sec:model}
\subsection{Normalization}\label{subsec:norm}
Let us first rewrite the SN equations (\ref{schrodinger})-(\ref{poisson}) in dimensionless form, using the analog of atomic units for the gravitational interaction. Thus, lengths are measured in units of the gravitational ``Bohr radius" $a_G =\hbar^2/(Gm^3)$, energy is measured in units of $E_G=m^5G^2/\hbar^2$ (the gravitational equivalent of the Hartree), and time in units of $t_G=\hbar/E_G$. To ensure conservation of the wavefunction norm, one also needs to normalize $\Psi$ to $a_G^{-3/2}$. The dimensionless SN equations then read as
\be
i \frac{\partial \Psi}{\partial t} = -\frac{1}{2} \Delta \Psi +V({\mathbf r},t) \Psi,
\label{schrodinger_norm}
\ee
\be
\Delta V = 4\pi |\Psi|^2,
\label{poisson_norm}
\ee
with the normalization condition $\int |\Psi|^2 d {\mathbf r}=1$. Notice that Eqs. (\ref{schrodinger_norm})-(\ref{poisson_norm}) are now free of all parameters. The evolution of the system is then entirely determined by its
initial condition. For instance, if the initial condition is spherically symmetric and Gaussian (as will be the case in the rest of this paper), the only relevant dimensionless parameter is the width of the initial Gaussian measured in units of $a_G$.

\subsection{Lagrangian approach}\label{subsec:lagrangian}
In this section, we will follow the derivation described in Refs. \cite{haas,manfredi_ep} in the context of atomic or condensed matter physics, where the relevant interaction is Coulombian rather than gravitational.

The SN equations (\ref{schrodinger_norm})-(\ref{poisson_norm}) can be written in a hydrodynamical form by using the Madelung transformation $\Psi = \sqrt{\rho}\exp(iS)$, where $\sqrt{\rho}$ is the amplitude and $S({\mathbf r},t)$ is the phase of the wavefunction \cite{manfredi_hydro}. The hydrodynamical continuity and momentum equations read as:
\begin{eqnarray}
\label{continuity}
\frac{\partial \rho}{\partial t} &+& \nabla\cdot\,(\rho{\bf u}) = 0 \,,\\
\label{momentum}
 \frac{\partial{\bf u}}{\partial t} &+& {\bf u}\cdot\nabla{\bf u}  = - \nabla V + \frac{1}{2} ~\nabla
\left(\frac{\nabla^2 \sqrt{\rho}}{\sqrt{\rho}} \right) ,
\end{eqnarray}
and the velocity is defined as the gradient of the phase, ${\mathbf u}=\nabla S$.

It can be shown that the above hydrodynamic equations (\ref{continuity})-(\ref{momentum}), together with Poisson's equation (\ref{poisson_norm}), can be derived from the following Lagrangian density ${\cal L}$ \cite{manfredi_ep}:
\be
{\cal L}(\rho,S,V) = \frac{\rho}{2}\left(\nabla S\right)^2 + \rho \frac{\partial S}{\partial t} +  \frac{\left(\nabla \rho\right)^2}{8\rho}
+\frac{(\nabla V)^2}{8\pi} +\rho V ~.
\label{lag_density}
\ee
So far, no approximation was made. The purpose is now to derive a set of evolution equations for a small number of macroscopic quantities that characterize the matter density profile.
With this aim in mind, let us assume that the system is spherically symmetric and that the density profile is Gaussian:
\begin{equation}
\label{gauss}
\rho({\mathbf r},t) = \frac{1}{\pi^{3/2} R^3(t)}\,\exp\left(- \frac{r^2}{R^2(t)}\right) \,,
\end{equation}
where $r = |{\mathbf r}|$ and $R(t)$ is the time-dependent size of the density.
For the above density profile, the exact solution of Poisson's equation (\ref{poisson_norm}) is
\begin{equation}
V({\mathbf r},t) = -\frac{1}{r} \,{\rm erf}\left(\frac{r}{R(t)}\right) \,,
\label{potential}
\end{equation}
where ${\rm erf}(x)$ is the error function.
In addition, the continuity equation (\ref{continuity}) is exactly solved by the following velocity field: ${\bf u} = (\dot R/R){\mathbf r}$,
which stems from the phase function $S = (\dot R/2R)r^2$. The dot denotes derivation with respect to time.

We can now compute the Lagrangian by plugging Eq. (\ref{gauss}) and the above solutions for $V$ and $S$ into Eq. (\ref{lag_density}), and integrating over all space, i.e., $L = -{2 \over 3}\int{\cal L}\,d{\bf r}$, where the multiplicative factor was introduced for convenience of notation. The result is
\be
L(R,\dot{R}) = \frac{\dot{R}^2}{2} - \frac{1}{2R^2} + \frac{C}{R}~,
\label{lag_au}
\ee
where $C=2/(3\sqrt{2\pi}).$
The corresponding equations of motion are obtained from the Euler-Lagrange equations
\be
\frac{d}{d t} \frac{\partial L}{\partial {\dot R}} -
 \frac{\partial L}{\partial R} =0~,
\ee
which yield:
\be
\frac{d^2 R}{d t^2} = \frac{1}{R^3} - \frac{C}{R^2}.
\label{eq_motion}
\ee

Equation (\ref{eq_motion}) is equivalent to the Hamiltonian equation of motion of a pointlike particle evolving in the external potential $U(R)=1/(2R^2) - C/R$ (see Fig. 1). The first term is repulsive and represents kinetic energy due to velocity dispersion (uncertainty principle), whereas the second term is attractive and represents self-gravity.

Note that this result was obtained from a rigorous development based on a Lagrangian variational principle. In particular, no assumptions of linear response were made in the derivation, so that Eq. (\ref{eq_motion}) can be used to extract information on the nonlinear regime of the time-dependent SN equations. { We also stress that the evolution obtained with the variational method is by construction unitary, since the trial Gaussian density [Eq. (\ref{gauss})] automatically satisfies: $\int\rho\,d{\bf r} = \int|\Psi|^2 d{\bf r} = 1$ for all times.}

\begin{figure}[ht]
\centering
\includegraphics[width=0.5\textwidth]{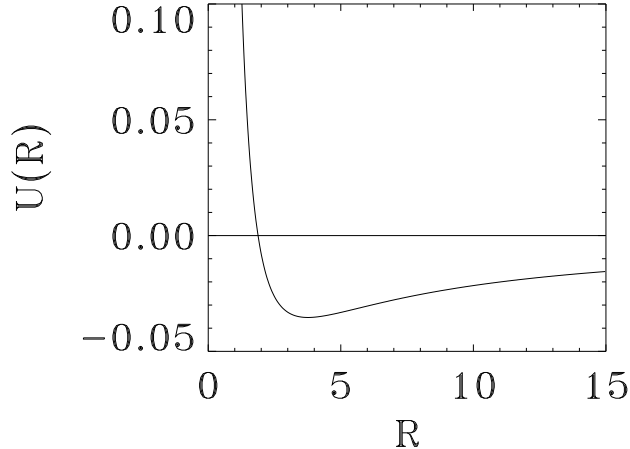}
\caption{Radial profile of the pseudo-potential $U(R)=1/(2R^2) - C/R$.} \label{fig:poten}
\end{figure}

\section{Results}\label{sec:results}
The pseudo-potential $U(R)$ is plotted in Fig. \ref{fig:poten}. It goes to infinity for $R \to 0$ and goes to zero as $R^{-1}$ for $R \to \infty$. It crosses the horizontal axis at a point $R_0$ such that $U(R_0)=0$ and has a single minimum at $R_1$, where $U'(R_1)=0$.
The values of these two points are easily determined and yield:
\be
R_0 = \frac{1}{2C}=\frac{3}{4}\sqrt{2\pi} \approx 1.88~;~~~~
R_1 = 2R_0=\frac{3}{2}\sqrt{2\pi} \approx 3.76.
\ee

\subsection{Ground state}\label{subsec:groundstate}
\begin{figure}[ht]
\centering
\includegraphics[width=0.5\textwidth]{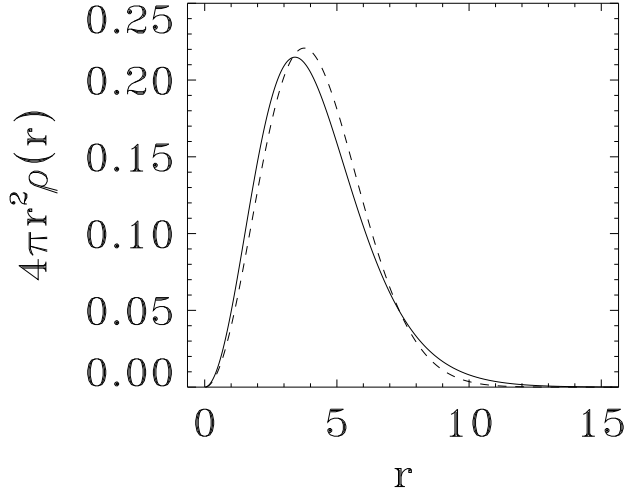}
\caption{Ground state density: numerically-computed profile (solid line) and Gaussian profile from Eq. (\ref{gauss}) with $R=R_1$ (dashed line).} \label{fig:density}
\end{figure}

The matter density profile in the ground state is given by Eq. (\ref{gauss}), with $R=R_1$, corresponding to the minimum of $U$. This profile is shown in Fig. \ref{fig:density} (dashed line), together with the ground state density obtained from a numerical solution of the stationary SN equations (solid line). The agreement is very good.

Stationary solutions of the SN equations must satisfy the virial theorem, which states that the potential energy (in absolute value) is twice the kinetic energy. We can verify that this is the case using the wavefunction $\Psi=\sqrt{\rho}$ from Eq. (\ref{gauss}) and the potential of Eq. (\ref{potential}).
The kinetic energy is
\be
K = \frac{1}{2}~\int\left(\frac{d\Psi}{dr}\right)^2 4\pi r^2dr =
\frac{3}{4 R^2},
\ee
whereas the potential energy yields:
\be
P = \frac{1}{2}~\int\rho V 4\pi r^2dr =
- \frac{1}{\sqrt{2\pi}R}.
\ee
It is readily checked that, when $R=R_1$, $|P|=2K=1/3\pi$, so that the virial theorem is satisfied.

The energy eigenvalue of the ground state (lowest energy state) has been computed numerically many times and an accepted value is $E_0=-0.163$ \cite{harrison,ruffini}.
{More accurate solutions may be obtained using the methods outlined in \cite{mansfield}}.
In our notation, $E_0=K+2P=-3K=-1/2\pi \approx -0.159$, which is rather close to the numerical value (the error is less than 3\%).

Finally, we note that, when $R=R_0$, one obtains $P=-K$ so that the total energy is zero.

\subsection{Dynamics}\label{subsec:dynamics}
One can linearize the equation of motion (\ref{eq_motion}) around the equilibrium, by writing $R=R_1 + \delta R$, where $\delta R(t)$ is a small perturbation. Substituting into Eq. (\ref{eq_motion}) and taking the Fourier transform in the time variable (i.e., assuming $\delta R(t) \sim e^{i\Omega t}$) yields the following oscillation frequency:
\be
|\Omega| = \sqrt{\frac{3}{R_1^4} - \frac{2C}{R_1^3}} = \frac{2}{9\pi} \approx 0.0707.
\label{omega}
\ee

A perturbation analysis of the full time-dependent SN equations was performed by Harrison et al. \cite{harrison}. The lowest oscillation frequency that these authors find (see Fig. 2 in Ref. \cite{harrison}) is close to $\Omega_{\rm Harr} = 0.035$, which is in very good agreement with Eq. (\ref{omega}) (the extra factor of 2 comes from the fact that Harrison et al. perturb the wavefunction instead of the density $|\Psi|^2$).

\begin{figure}[ht]
\centering
\includegraphics[width=0.7\textwidth]{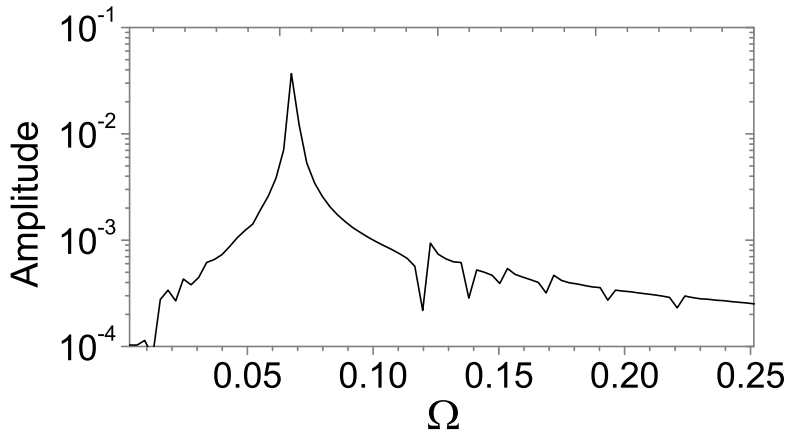}
\caption{Frequency spectrum of $\sqrt{\langle r^2 (t) \rangle}$ for small oscillations around the ground-state equilibrium, obtained from the full time-dependent SN equations.} \label{fig:spectrum}
\end{figure}

In order to check this result, we solved the spherically symmetric time-dependent SN equations, using a second-order Crank-Nicolson method with centred differences for the spatial differentiation. The initial condition is the exact ground state computed numerically (solid curve in Fig. \ref{fig:density}), to which a very small perturbation was added. The root mean square of the radius $\sqrt{\langle r^2 (t) \rangle}$ is then computed using the standard quantum average. Its frequency spectrum is shown in Fig. \ref{fig:spectrum} and displays a clear peak around $\Omega = 0.067$, which is again very close to Eq. (\ref{omega}).

\begin{figure}[ht]
\centering
\includegraphics[width=0.7\textwidth]{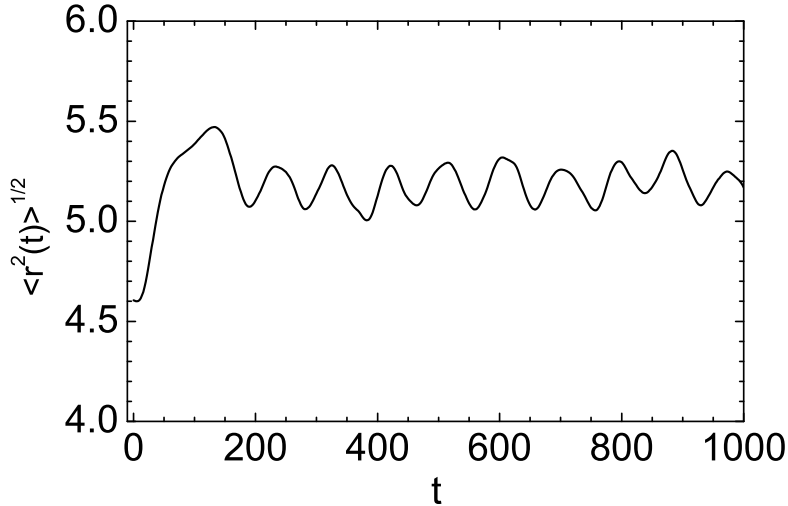}
\caption{Evolution of the mean square radius $\sqrt{\langle r^2(t) \rangle}$ for the full time-dependent SN equations with a Gaussian initial condition given by Eq. (\ref{gauss}) with $R(0)=R_1$.} \label{fig:radius}
\end{figure}

For an initial condition that is slightly farther from the exact ground state, we expect the SN equations to display some nonlinear effects.
This is apparent from Fig. \ref{fig:radius}, where the Gaussian profile given by Eq. (\ref{gauss}) with $R(0)=R_1$ (dashed line in Fig. 2) was used as an initial condition.
The time history of $\sqrt{\langle r^2 (t) \rangle}$ clearly shows some nonlinear oscillations, although their frequency is still close to the linear estimate given by Eq. (\ref{omega}).

\begin{figure}[ht]
\centering
\includegraphics[width=0.45\textwidth]{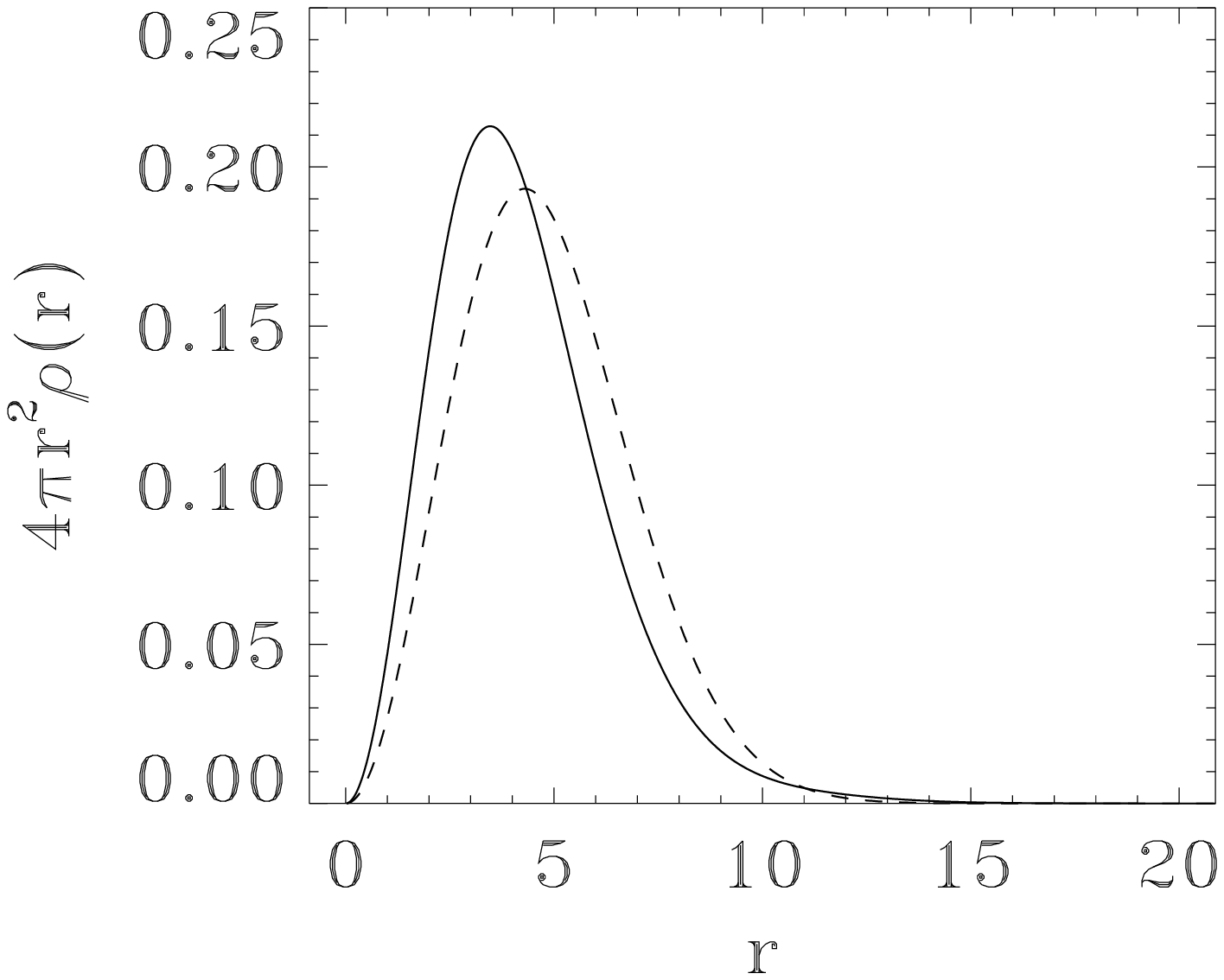}
\includegraphics[width=0.47\textwidth]{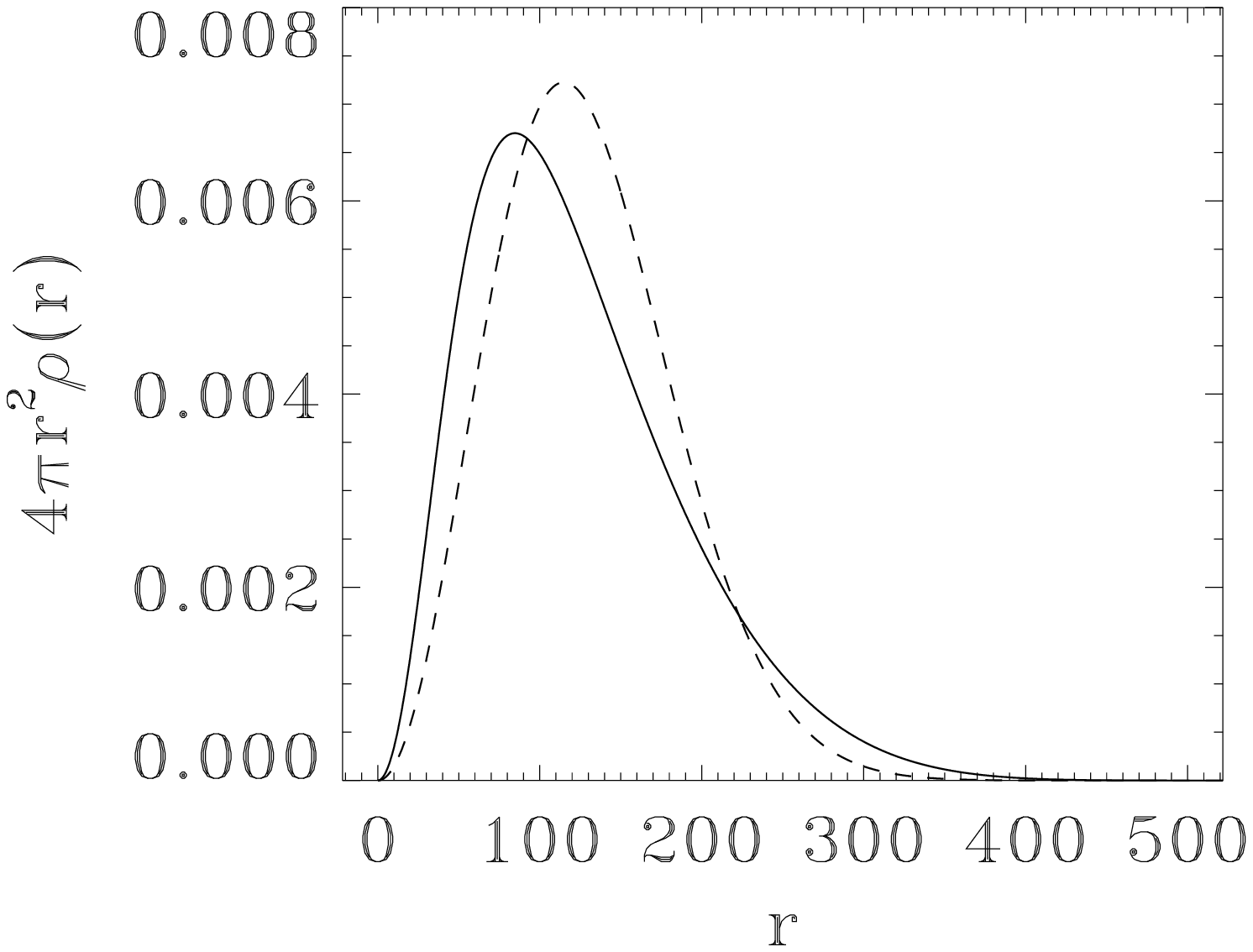}
\caption{Numerically-computed density (solid line) and corresponding Gaussian density with same width (dashed line). Left panel: same case as in Fig. \ref{fig:radius} at time $t=500$; Right panel: expanding solution with initial width $R(0)=0.3R_1$, plotted at time $t=200$.} \label{fig:check}
\end{figure}

{ It is also useful to monitor the evolution of the density $\rho({\mathbf r},t)$ in order to check that it stays sufficiently close to a Gaussian function, which is required for the validity of the variational approach. This is done in Fig. \ref{fig:check}, where we plot the mass density obtained from numerical simulations of the full SN equations (solid lines) and compare it to a Gaussian density [Eq. (\ref{gauss})] with same width (dashed lines).
The left panel refers to the same evolution as in Fig. \ref{fig:radius} at $t=500$. The right panel refers to a case where the density is initially localised near the origin, so that $R(0) < R_1$. In this case, the system expands almost freely and at $t=200$ (corresponding to the plot of Fig. \ref{fig:check}) it has attained a considerable size. In both cases, the numerically-computed density is reasonably close to a Gaussian profile, thus strengthening our confidence in the present variational approach. We also stress that for similar problems involving the Coulomb interaction and rather strong nonlinearities (quartic confinement), the variational procedure appeared to work rather well \cite{haas}.
}

\begin{figure}[ht]
\centering
\includegraphics[width=0.6\textwidth]{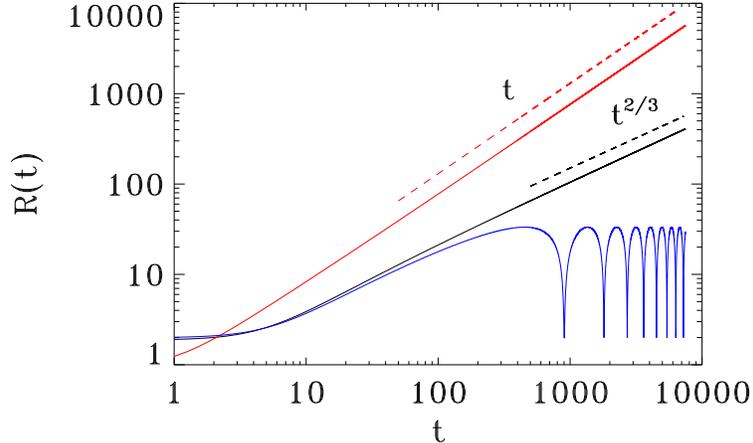}
\caption{Solutions of the equations of motion (\ref{eq_motion}) for three initial conditions: $R(0)<R_0$ (red solid line), $R(0)>R_0$ (blue line), and $R(0)=R_0$ (black solid line). The dashed straight lines represent the curves $R \sim t$ (red) and $R \sim t^{2/3}$ (black).} \label{fig:expansion}
\end{figure}

Finally, we consider the long-time solutions of Eq. (\ref{eq_motion}). From the shape of the pseudo-potential $U(R)$ (Fig. 1), it is clear that three different regimes are possible for an initial condition $R(0)>0$, $\dot R(0)=0$:
\begin{itemize}
\item
If $R(0)<R_0$, the total energy is positive, i.e., kinetic energy dominates over gravitational energy. In this case, the wave packet expands indefinitely. Although the expansion is slowed down initially by the gravitational attraction, the asymptotic evolution ($t \to \infty$) is that of a free particle, i.e., $R \sim t$.
\item
If $R(0)>R_0$, the total energy is negative, i.e., gravitational energy dominates over kinetic energy. The wave packet oscillates at a nonlinear frequency that can in principle be computed from the expression of $U(R)$ (it reduces to the linear frequency $\Omega$ when $R(0) \approx R_1$).
\item
If $R(0)=R_0$, the total energy is exactly zero. The wave packet still expands, but at a rate slower than $R \sim t$. The first term on the right-hand side of Eq. (\ref{eq_motion}) becomes negligible for long times. Matching the remaining two terms shows that the expansion should go like $R \sim t^{2/3}$.
\end{itemize}
The three regimes described above are neatly reproduced in numerical simulations of Eq. (\ref{eq_motion}), shown in Fig. \ref{fig:expansion}. It is also worth to note that in cosmology these regimes correspond respectively to an open, closed, and Einstein-de Sitter universe (which expands as $t^{2/3}$).

Of course, the full evolution of the wavefunction according to the SN equations can be much richer than this simple picture. For large masses, the wavepacket can break down into two parts, with some mass being ejected to infinity while the rest remains confined \cite{giulini2011,vanmeter}. This behaviour cannot be captured by our { variational} approach, which postulates that the density remains close to a Gaussian profile for all times.
{ One could nevertheless extend the present model by considering more complicated trial functions involving more than one variational parameter. This would result in a set of coupled nonlinear differential equations that generalize Eq. (\ref{eq_motion}).}

\section{Discussion}\label{sec:discussion}
The main interest of the method outlined in this paper is that
it relies on a rigorous development based on a variational principle, while at the same time yielding results that are simple and intuitive.
The very shape of the pseudo-potential $U(R)$ (Fig. \ref{fig:poten}) informs us on the type of motions that are to be expected.
For instance, it is clear that wavepacket dispersion occurs if $R<R_0$, whereas it is inhibited if $R>R_0$. Further, if $R> R_1 \approx 3.76 \hbar^2/(Gm^3)$ the wavepacket should start to contract right from the beginning of the evolution (for clarity, we restore dimensional units in this section).

Now, we want to compare the above estimations with the numerical results of
Giulini et al. \cite{giulini2011}, who considered a system of initial size $R=0.707 \rm \mu m$ ($a= 0.5 \rm \mu m$ in their notation). They observed a contracting wavepacket for masses greater than $7 \times 10^9$amu (atomic mass units), which is rather close to the value $m=5.74 \times 10^9$amu predicted by our formula $R m^3 = 3.76~ \hbar^2/G$. The results of other simulations \cite{vanmeter} are also consistent with these findings.

\begin{figure}[ht]
\centering
\includegraphics[width=0.6\textwidth]{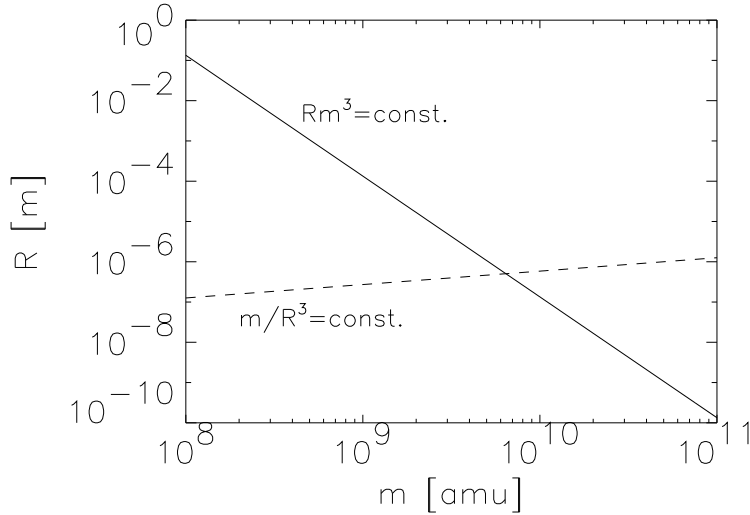}
\caption{Mass-radius diagram. Gravitationally induced effects should be important in the region above the solid line, which corresponds to the curve $R m^3 > 3.76~ \hbar^2/G$. The dashed curve corresponds to a constant density: $m{\rm [amu]}/R^3= 5\times 10^{28} \rm m^{-3}$ (typical solid-state density).} \label{fig:mass}
\end{figure}

It is clear that the importance of self-gravitational effects in the SN equations depends on both the size $R$ and the mass $m$ of the object under consideration. Therefore, it is useful to plot a mass-radius diagram on a log-log scale (Fig. \ref{fig:mass}), where these two quantities appear explicitly.
Gravitational effects should play a significant role for objects that fall in the region above the curve defined by $R m^3 = {\rm const.} = 3.76~ \hbar^2/G$ (solid line).

Experiments aimed at detecting the role of gravity on quantum decoherence will probably involve studying the interference fringes of solid-state mesoscopic objects, which should be light enough for quantum coherence to be observable but also heavy enough for gravitational effects to play a measurable role. Interferometry experiments on small silica spheres \cite{romero} and gold clusters \cite{nimmrichter} are possible candidates for such studies.

In order to fix ideas, let us focus on the case of gold or other metal clusters, for which the number density is typically $n_{\rm gold} \approx 5\times 10^{28} \rm m^{-3}$. The dashed line on Fig. \ref{fig:mass} represents the curve at constant density $m {\rm [amu]} /R^3 = n_{\rm gold}$.
The intersection of the dashed line with the the solid line $R m^3 = 3.76 \hbar^2/G$ yields the minimum mass and radius that gold clusters should possess for gravitational effects to play a significant role. This turns out to be of the order of a few microns in size and about $5 \times 10^9$ in atomic mass units. The same calculation performed for other metal clusters or the silica spheres mentioned above yields similar results.

The experimental challenge will be to perform quantum interference experiments on such massive objects and to control other non-gravitational sources of decoherence. In practice, one may perform different experiments for increasing values of the cluster mass, thus moving from left to right on the dashed line in Fig. \ref{fig:mass}. When crossing the solid line, gravitational effects should be detected, perhaps as a reduction in the contrast of the interference fringes.

\paragraph{Acknowledgments.}
F.H. thanks the CNPq (Conselho Nacional de Desenvolvimento Cient\'{\i}fico e Tecnol\'ogico) for partial financial support.

\end{document}